\pdfoutput=1

\documentclass[11pt]{article}

\usepackage[final]{acl}

\usepackage{times}
\usepackage{latexsym}

\usepackage[T1]{fontenc}

\usepackage[utf8]{inputenc}

\usepackage{microtype}

\usepackage{inconsolata}

\usepackage{graphicx}
\usepackage{array}
\usepackage{multirow}
\usepackage{amsmath}
\usepackage{makecell}

\title{Audio-Based Linguistic Feature Extraction for Enhancing Multi-lingual and Low-Resource Text-to-Speech}

 \author{Youngjae Kim$^{1}$, Yejin Jeon$^{1}$, Gary Geunbae Lee$^{1,2}$ \\
         $^{1}$ Graduate School of Artificial Intelligence, POSTECH, Pohang, South Korea \\ 
         $^{2}$ Computer Science and Engineering, POSTECH, Pohang, South Korea \\
         \texttt{yj122198@postech.ac.kr}, \texttt{jeonyj0612@postech.ac.kr}, \texttt{gblee@postech.ac.kr}}

\begin{document}
\maketitle
\begin{abstract} 
The difficulty of acquiring abundant, high-quality data, especially in multi-lingual contexts, has sparked interest in addressing low-resource scenarios. Moreover, current literature rely on fixed expressions from language IDs, which results in the inadequate learning of language representations, and the failure to generate speech in unseen languages. To address these challenges, we propose a novel method that directly extracts linguistic features from audio input while effectively filtering out miscellaneous acoustic information including speaker-specific attributes like timbre. Subjective and objective evaluations affirm the effectiveness of our approach for multi-lingual text-to-speech, and highlight its superiority in low-resource transfer learning for previously unseen language.

\end{abstract}

\section{Introduction}
Text-to-speech (TTS) models have achieved remarkable advancements in generating human-like speech with a high degree of clarity and naturalness \cite{FastSpeech2, VITS, Naturalspeech}. However, these models necessitate an immense volume of high-quality audio data for training to produce such high-quality synthetic speech. This challenge becomes even more pronounced when extending a mono-lingual TTS model to multi-lingual setting, as each additional language requires a proportional increase in training data. To mitigate this data-intensive challenge, \citet{TransferTTS_multilingual, phat_do_2022, phat_do_2023} have employed transfer learning techniques to synthesize speech for low-resource languages. Nevertheless, a substantial amount of source language corpora is still necessary to generate the linguistic knowledge that can be reused for speech synthesis in low-resource languages. 

\begin{figure}[t]
    \centering
    \includegraphics[width=\columnwidth]{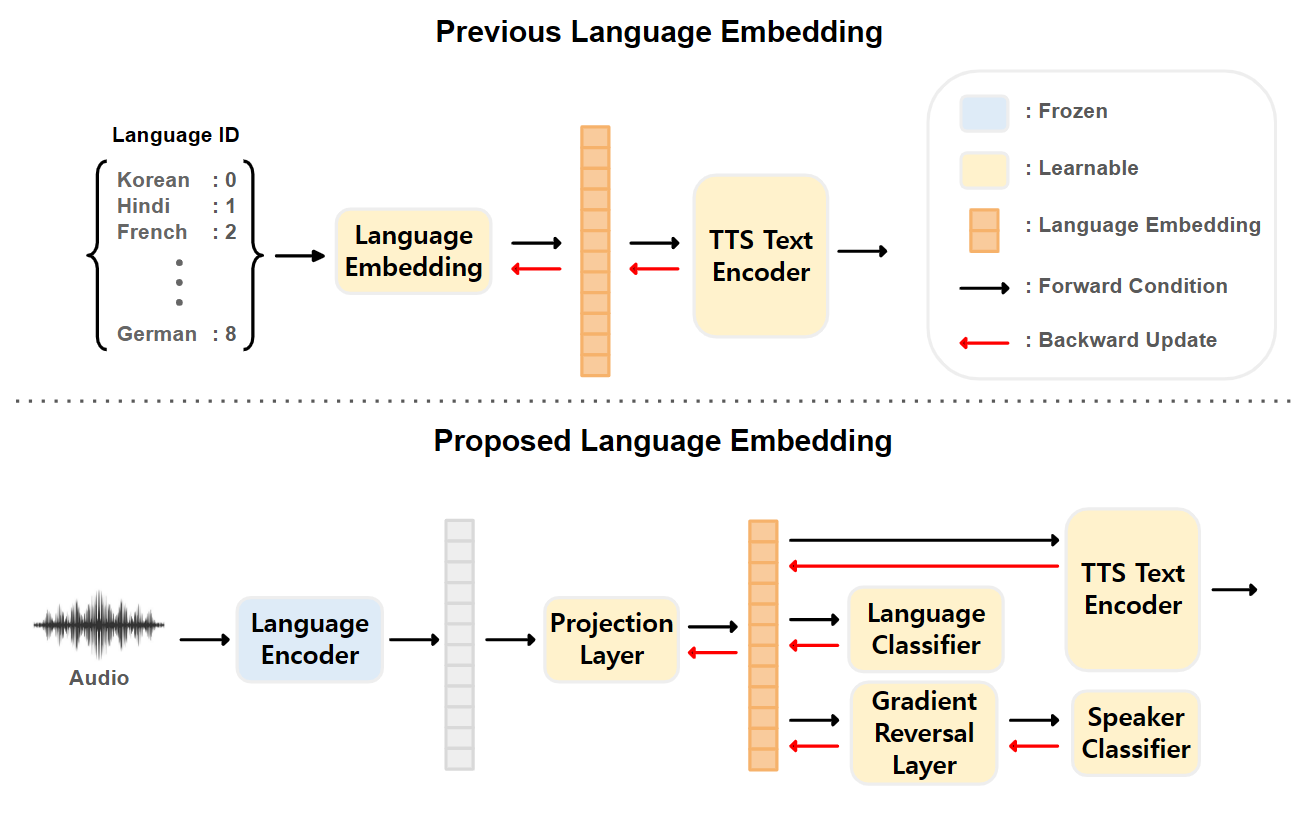}
    \caption{Conceptual visualization comparing prior multi-lingual methodologies that utilize language IDs for language representations (upper) with the proposed methodology (lower). The architecture after the text encoder follows \citet{VITS}, which we omit for brevity.}
    \label{fig:architecture}
\end{figure}

The predominant method employed in current multi-lingual TTS systems involve the use of unique language IDs assigned to each language \cite{casanova2022yourtts, sane_tts, zhang2019learning, Meta_multilingual}. In other words, during both the training and inference phases, the model requires the target language's unique ID as additional input to generate speech in that language. However, this approach presents two significant limitations. First, it is incapable of generating speech for unseen languages that the model has not been trained on. To synthesize audio in an unseen language, the TTS model must be retrained from scratch with a new language ID specifically assigned to the new language, rendering the expansion to additional languages cumbersome and inefficient. Second, merely representing each language with its corresponding fixed label fails to capture the diverse and intricate features inherent to each language. As a result, this limits the model's ability to accurately generate linguistic nuances within synthetic speech.

To address these challenges, we introduce a novel approach that directly learns linguistic features from audio input instead of relying on fixed expressions from language IDs. This innovative method facilitates the extraction and generation of representations with richer linguistic information. Furthermore, our approach enables the synthesis of speech in unseen language with minimal fine-tuning, requiring as little as ten minutes of data from the target language. Remarkably, not only does this method generate high-quality speech in unseen languages, but also improves speech synthesis for seen languages. Comprehensive evaluations confirm that our method achieves precise pronunciation in both seen and unseen languages, underscoring its versatility and effectiveness.

\section{Methods}

\subsection{Preliminaries}
The backbone TTS architecture adopted in this research is a non-autoregressive conditional variational autoencoder VITS \cite{VITS}. Initially, VITS processes an input text sentence $S$ to generate the corresponding audio using a Transformer-based \cite{transformers} text encoder $E_{text}$, a stochastic duration predictor, and a decoder. While the primary focus of this study is multi-lingual TTS, the VITS framework necessitates further refinement to incorporate speaker-specific information, given the absence of datasets featuring individuals proficient in multiple languages. Consequently, the conventional feed forward layer within the transformer encoder is replaced with a structured block comprising a linear layer, a kernel-based CNN layer, and another linear layer. The kernel-based CNN layer especially, is responsible for the integration of the speaker representation, which is derived by passing audio input $a$ into a reference encoder composed of 2D convolution layers and a GRU layer, with the latent representations of the textual input. Interested readers are encouraged to refer to \cite{SC-CNN}, as speaker integration extends beyond the immediate scope of multi-lingual TTS research. 
\subsection{Audio-Based Language Encoder}
This subsection introduces a novel approach that can capture nuanced linguistic features directly from acoustic input, and generate speech in unseen languages. The proposed architecture is shown in the lower part of Figure ~\ref{fig:architecture}.

Initially, to process audio input $a_{input}$, we utilize a pretrained language encoder $E_{lang}$ that is trained using a language identification task by Speechbrain \cite{speechbrain}.\footnote{\url{https://huggingface.co/speechbrain/lang-id-voxlingua107-ecapa}} Specifically, $E_{lang}$ is a modified Time Delay Neural Network (TDNN) \cite{ECAPA-TDNN} x-vector architecture that leverages Squeeze-Excitation Res2Blocks. The output from the pretrained $E_{lang}$ results in intermediate language representation of $z_{lang}$. 

From this acoustic representation, in order to enhance linguistic properties, $z_{lang}$ is fed through an additional 1D convolution projection layer $P$, which generates a new representation of $h_{lang}$. Subsequently, we utilize a language classifier to predict the language label of $h_{lang}$ using a softmax function. However, due to the diverse acoustic features present in the original acoustic input $a_{input}$, such as those related to speaker voice and gender, these extraneous attributes must also be filtered out. To do so, a speaker classifier is additionally incorporated, and a gradient reversal layer $R$ is utilized to facilitate speaker adversarial training (SAT). Thus, the holistic loss function $L_{le}$ for optimizing the language embedding $h_{lang}$ can be mathematically derived as follows, where $y_{lang}$ and $y_{spk}$ represent the the ground truth labels for language and speaker, respectively.
\begin{align}
\label{eq:lang_loss}
  L_{lang} &= -\sum \log p(y_{lang} | h_{lang}) \\
\label{eq:spk_loss}
  L_{spk} &= -\sum \log p(y_{spk} | R(h_{lang})) \\
\label{eq:total_loss}
  L_{le} &= L_{lang} + L_{spk} 
\end{align}

\subsection{Training Protocol}

The proposed model is trained using a two-stage protocol. In the first stage, we train the model with multi-lingual data, updating all parameters from scratch except the pretrained language encoder $E_{lang}$. In the second stage, the model is finetuned on a target low-resource language, and only the parameters following the language embedding $h_{lang}$ is optimized. In other words, just the text encoder, stochastic duration predictor, and the decoder of the backbone VITS are updated. By freezing all parameters up to the projection layer that generates $h_{lang}$, we ensure that the limited data of the target language does not excessively influence the previously pretrained components. This preserves the linguistic knowledge learned from multi-lingual data.

In both stages, $h_{lang}$ is integrated into the backbone VITS with SC-CNN. Alongside the language embedding loss $L_{le}$, we utilize additional loss functions from VITS: mel spectrogram reconstruction loss $L_{recon}$, KL-divergence loss $L_{kl}$, adversarial loss $L_{adv}$, feature matching loss $L_{fm}$, and duration loss $L_{dur}$. The overall loss functions for multi-lingual pretraining and low-resource finetuning settings are defined as follows:
\begin{equation}
    \begin{aligned}
        L_{ml} &= L_{recon} + L_{kl} + L_{adv} \\
               &\quad + L_{fm} + L_{dur} + L_{le}
    \end{aligned}
\end{equation}
\begin{equation}
        L_{lr} = L_{recon} + L_{kl} + L_{adv} + L_{fm} + L_{dur}
\end{equation}


\begin{table*}[t]
\centering
\renewcommand{\arraystretch}{1.4}
\setlength{\tabcolsep}{4pt}
\small
\begin{tabular*}{\textwidth}{@{\extracolsep{\fill}} l*{9}{c}@{}}
\hline
 & German & French & Italian & Korean & Hindi & Polish & Russian & Spanish & Ukrainian \\
\hline
Ground Truth & 2.10 & 4.02 & 3.06 & 6.55 & 13.02 & 2.62 & 3.34 & 2.49 & 4.09 \\
Baseline & 4.99 & 9.32 & 5.92 & 9.69 & 20.46 & 6.06 & 7.13 & 7.14 & 7.89 \\
Proposed (w/o SAT) & 4.61 & 8.33 & 5.51 & 9.01 & 18.48 & 4.85 & 6.53 & 5.88 & 7.61 \\
Proposed (with SAT) & \textbf{3.75} & \textbf{7.17} & \textbf{4.43} & \textbf{8.80} & \textbf{18.33} & \textbf{4.56} & \textbf{5.49} & \textbf{4.87} & \textbf{6.45} \\
\hline
\end{tabular*}
\caption{Results of pretrained multi-lingual TTS, with each language evaluated using CER (\%). Lower CER scores indicate better performance.}
\label{tab:multilingual_cer}
\end{table*}

\section{Experiments}

In this section, we provide a detailed description of the datasets, their preprocessing procedures, and the experimental settings required for both multi-lingual and low-resource settings.

\label{sec:dataset}
\subsection{Datasets}

Different datasets were utilized for each training stage of the model. In the initial multi-lingual pretraining phase, nine languages were used: Korean, German, French, Italian, Hindi, Spanish, Russian, Ukrainian, and Polish. Due to the challenge of evaluating performance with genuinely low-resource languages, English was assumed to be the target low-resource language for fine-tuning.

Since there is no unified dataset for these languages, we compiled various open-source datasets: the AI Hub \cite{dataset_ai_hub} dataset for Korean, the HUI-Audio-Corpus-German \cite{dataset_german} for German, the M-AILABS \cite{dataset_m-ailabs} dataset for Spanish, French, Italian, Russian, Polish, and Ukrainian, the LIMMITS24 \cite{dataset_limmits24} dataset for Hindi, and the LJSpeech \cite{dataset_ljspeech} dataset for English.

All text data were preprocessed using International Phonetic Alphabet (IPA) token, in line with previous TTS systems \cite{VITS, TransferTTS_multilingual, SC-CNN}, and audio data was resampled to 22.05 kHz. For multi-lingual training, we extracted 10 hours of training data per language and set aside 500 evaluation sentences per language. For the low-resource finetuning stage, we used 10 minutes and 1 hour of data from the English dataset, and prepared 500 evaluation sentences.

\subsection{Multi-lingual Experimental Settings} 

We used four Nvidia A5000 GPUs with a batch size of 32 to pretrain our model on multi-lingual data for 400K steps. To validate the performance of our pretrained model, we compared it against a MM-TTS \cite{mm-tts} model that used language IDs to represent individual languages. This baseline model was pretrained using the same multi-lingual data under the same experimental conditions as our pretrained model, utilizing the same type and number of GPUs, batch size, and training steps. Other configurations not explicitly mentioned followed the settings from original VITS setup.

\subsection{Low-resource Experimental Settings} 
For low-resource setting, we performed transfer learning, using the model trained in the multi-lingual setting as the pretrained model. For comparison, we used the original VITS model and conducted the experiments under the same conditions. The models were trained on four NVIDIA A5000 GPUs with a batch size of 16 for 30K steps.

\section{Results and Discussion}
For evaluation, we employed Mean Opinion Score (MOS) and Character Error Rate (CER) as metrics. The MOS test was conducted to assess the quality of the synthesized speech. Fifteen evaluators recruited from the Amazon Mechanical Turk platform listened to the synthesized audio samples and rated their naturalness on a Likert scale from 1 to 5. However, due to the difficulty of recruiting multilingual experts for the MOS evaluation, we employed NORESQA-MOS~\cite{noresqa}, a MOS prediction model, for the multilingual evaluation. Additionally, CER was measured to evaluate the intelligibility of the synthesized speech. We utilized pretrained speech recognition model\footnote{\url{https://github.com/openai/whisper}}: OpenAI's Whisper ~\cite{whisper} to transcribe the synthesized audio and calculate CER.

\subsection{Multi-lingual Results} 
As shown in Table~\ref{tab:multilingual_cer}, we conducted a multi-lingual evaluation using CER. Our proposed model demonstrated superior performance over the baseline model across all languages. Additionally, we performed MOS prediction using NORESQA-MOS. The predicted scores are as follows: the baseline model had an average score of 4.32, the proposed method without SAT scored 4.41, and the proposed method with SAT scored 4.44. Detailed scores are available in the Appendix ~\ref{sec:appendix}. These results demonstrate that our method of extracting language embedding directly from audio is able to effectively capture language representations compared to previous methods relying on language IDs.

\subsection{Low-resource Result} 

We utilize both CER and MOS metrics to evaluate performance in the low-resource setting. As shown in Table ~\ref{tab:finetune_result}, our method showed strong performance in both metrics. Specifically, while the original VITS model does not perform adequately in the low-resource setting (high CER and low MOS scores), our proposed model with SAT exhibited robust performance. In some instances, our methodology even surpasses the ground truth CER results. This illustrates our method's capability to learn more nuanced and effective language representations, which are effective even for previously unseen language. 

\begin{table}[t]
\centering
\renewcommand{\arraystretch}{1.4}
\small
\begin{tabular}{llcc}
\hline
\multicolumn{2}{c}{\textbf{Model}} & \textbf{CER} ($\downarrow$) & \textbf{MOS} ($\uparrow$) \\ \hline
\multicolumn{2}{c}{Ground Truth} & 2.31 & 4.14 ± 0.12 \\ \hline
\multicolumn{1}{c|}{} & VITS & 3.88 & 2.69 ± 0.21 \\
\multicolumn{1}{c|}{\textbf{10m}} & Proposed (w/o SAT) & 2.32 & 3.87 ± 0.13 \\
\multicolumn{1}{c|}{} & Proposed (with SAT) & \textbf{2.14} & \textbf{3.97 ± 0.12} \\ \hline
\multicolumn{1}{c|}{} & VITS & 2.19 & 3.47 ± 0.14 \\
\multicolumn{1}{c|}{\textbf{1h}} & Proposed (w/o SAT) & 1.96 & 4.02 ± 0.12 \\
\multicolumn{1}{c|}{} & Proposed (with SAT) & \textbf{1.84} & \textbf{4.11 ± 0.13} \\ \hline
\end{tabular}
\caption{CER (\%) and MOS scores for low-resource setting with 95\% confidence intervals.}
\label{tab:finetune_result}
\end{table}

\subsection{Ablation Study}
To thoroughly validate our methodology, we conducted two additional experiments. First, we compared the performance with and without speaker adversarial training (SAT) to completely remove speaker information from the language embedding extracted from the audio. As shown in Tables ~\ref{tab:multilingual_cer} and ~\ref{tab:finetune_result}, performance improved in all cases when SAT was included. This confirms that removing speaker information from the language embedding allows for better conveyance of language representations.

\begin{figure}
    \centering
    \includegraphics[width=\columnwidth]{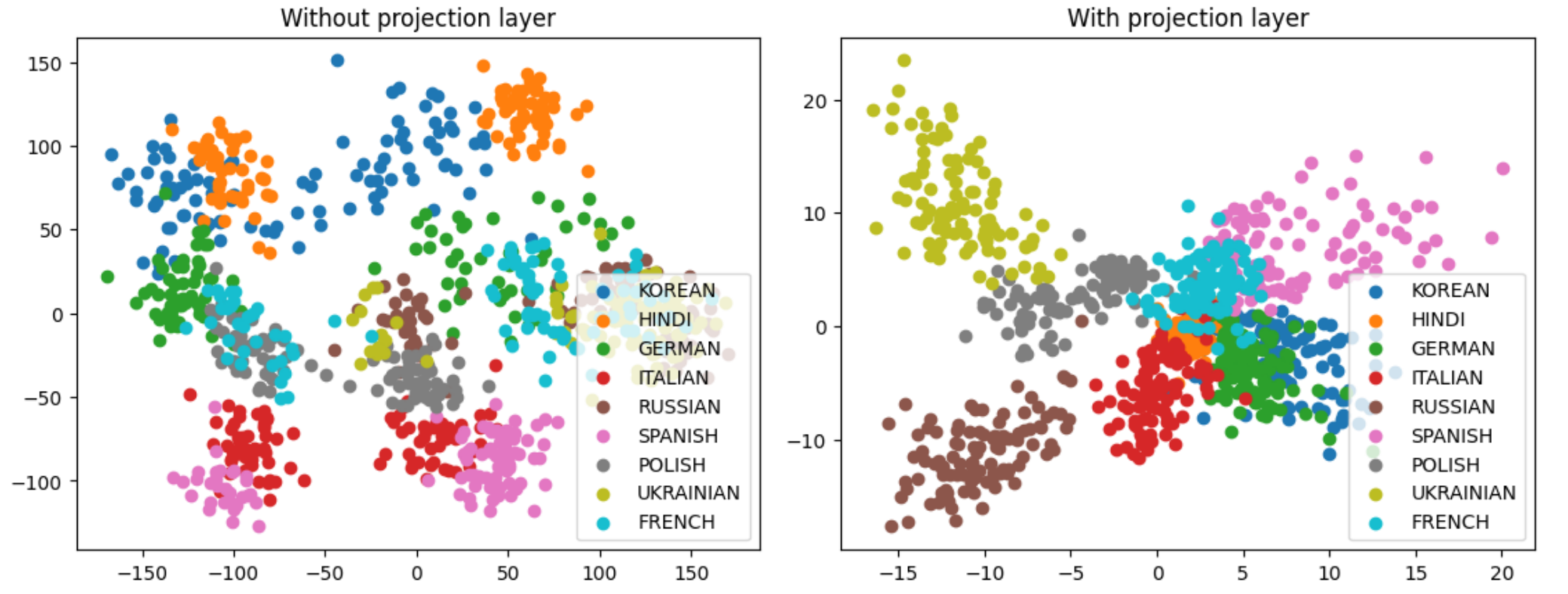}
    \caption{Distribution of language embeddings with and without the projection layer. Visualizations were conducted using 2D PCA. Utilization of an addition projection layer results in distinct language-specific clusters.}
    \label{fig:language_embedding}
\end{figure}

Moreover, we examined the distribution of the language embedding with and without the projection layer added to the pretrained language encoder. Figure ~\ref{fig:language_embedding} shows the 2D PCA visualization of language embedding extracted under both conditions. Without the projection layer, the language distributions varied significantly depending on gender or speaker, indicating that the language embedding was influenced by these characteristics. However, the additional projection layer and training with classifiers enabled the language embedding to become more consistent and clustered according to language. This demonstrates that the projection layer enhances the consistency of language embedding, thereby improving the clarity of language distinctions and enhancing the model's performance.

\section{Conclusion}
In this paper, we proposed a method for multi-lingual TTS  that directly extracts language embedding from audio rather than using fixed expressions from language IDs. We enhanced the pretrained language encoder with a projection layer to better capture language representations and incorporated speaker adversarial training to remove extraneous speaker information from the language embedding. Our experimental results demonstrate that this straightforward approach is highly effective for multi-lingual TTS. Furthermore, we were able to observe significant improvements compared to previous methodology. Whereas previous language ID approach was unable to generate speech for unseen languages, our proposed approach allows for the generation of high-quality speech for an unseen language by finetuning the model using just 10 minutes of additional data. This promising outcome suggests that our method could significantly enhance the accessibility and quality of TTS across multi-lingual and low-resource settings.

\section{Limitations}


In order to comprehensively evaluate the efficacy of our proposed method, as discussed in Section ~\ref{sec:dataset}, we considered English as a low-resource language for testing purposes. It is essential to extend this evaluation to actual low-resource languages in future studies, as our current approach serves as a proof of concept.
Furthermore, while we utilized a language classifier to aid in linguistic knowledge learning, employing alternative methods such as unsupervised learning could potentially offer a more robust approach. Therefore, future research will explore diverse learning methods for extracting language embeddings from audio data.

\section{Ethical Considerations}
The target task of speech synthesis can potentially be used to create artificial speech without someone's consent. As such, we place importance on conducting responsible and ethical research.

\section*{Acknowledgments}
This work was supported by Smart HealthCare Program(www.kipot.or.kr) funded by the Korean National Police Agency(KNPA, Korea) [Project Name: Development of an Intelligent Big Data Integrated Platform for Police Officers’ Personalized Healthcare / Project Number: 220222M01] and also by the MSIT(Ministry of Science and ICT), Korea, under the ITRC(Information Technology Research Center) support program(IITP-2024-RS-2024-00437866) supervised by the IITP(Institute for Information \& Communications Technology Planning \& Evaluation)

\bibliography{main}

\begin{thebibliography}{22}
\providecommand{\natexlab}[1]{#1}

\bibitem[{Casanova et~al.(2022)Casanova, Weber, Shulby, Junior, G{\"o}lge, and Ponti}]{casanova2022yourtts}
Edresson Casanova, Julian Weber, Christopher~D Shulby, Arnaldo~Candido Junior, Eren G{\"o}lge, and Moacir~A Ponti. 2022.
\newblock Yourtts: Towards zero-shot multi-speaker tts and zero-shot voice conversion for everyone.
\newblock In \emph{International Conference on Machine Learning}, pages 2709--2720. PMLR.

\bibitem[{Cho et~al.(2022)Cho, Jung, Lee, and Woo}]{sane_tts}
Hyunjae Cho, Wonbin Jung, Junhyeok Lee, and Sang~Hoon Woo. 2022.
\newblock Sane-tts: Stable and natural end-to-end multilingual text-to-speech.
\newblock In \emph{International Conference on Machine Learning}.

\bibitem[{Desplanques et~al.(2020)Desplanques, Thienpondt, and Demuynck}]{ECAPA-TDNN}
Brecht Desplanques, Jenthe Thienpondt, and Kris Demuynck. 2020.
\newblock {ECAPA-TDNN: Emphasized Channel Attention, propagation and aggregation in TDNN based speaker verification}.
\newblock In \emph{Interspeech 2020}, pages 3830--3834.

\bibitem[{Do et~al.(2022)Do, Coler, Dijkstra, and Klabbers}]{phat_do_2022}
Phat Do, Matt Coler, J.E. Dijkstra, and Esther Klabbers. 2022.
\newblock Text-to-speech for under-resourced languages: Phoneme mapping and source language selection in transfer learning.
\newblock In \emph{Proceedings of the the 1st Annual Meeting of the ELRA/ISCA Special Interest Group on Under-Resourced Languages}, pages 16--22. European Language Resources Association (ELRA).

\bibitem[{Do et~al.(2023)Do, Coler, Dijkstra, and Klabbers}]{phat_do_2023}
Phat Do, Matt Coler, Jelske Dijkstra, and Esther Klabbers. 2023.
\newblock \href {https://doi.org/10.21437/SSW.2023-4} {{Strategies in Transfer Learning for Low-Resource Speech Synthesis: Phone Mapping, Features Input, and Source Language Selection}}.
\newblock In \emph{Proc. 12th ISCA Speech Synthesis Workshop (SSW2023)}, pages 21--26.

\bibitem[{Ito and Johnson(2017)}]{dataset_ljspeech}
Keith Ito and Linda Johnson. 2017.
\newblock The lj speech dataset.
\newblock \url{https://keithito.com/LJ-Speech-Dataset/}.

\bibitem[{Jeon et~al.(2024)Jeon, Kim, and Lee}]{mm-tts}
Yejin Jeon, Youngjae Kim, and Gary~Geunbae Lee. 2024.
\newblock \href {https://sigport.org/documents/leveraging-effective-language-and-speaker-conditioning-indic-tts-limmits-2024-challenge-0} {Leveraging effective language and speaker conditioning in indic tts for limmits 2024 challenge}.

\bibitem[{Jeong et~al.(2024)Jeong, Kim, Choi, Yoon, Jang, and Kim}]{TransferTTS_multilingual}
Myeonghun Jeong, Minchan Kim, Byoung~Jin Choi, Jaesam Yoon, Won Jang, and Nam~Soo Kim. 2024.
\newblock \href {https://doi.org/10.1109/TASLP.2024.3364085} {Transfer learning for low-resource, multi-lingual, and zero-shot multi-speaker text-to-speech}.
\newblock \emph{IEEE/ACM Trans. Audio, Speech and Lang. Proc.}, 32:1519–1530.

\bibitem[{Kim et~al.(2021)Kim, Kong, and Son}]{VITS}
Jaehyeon Kim, Jungil Kong, and Juhee Son. 2021.
\newblock Conditional variational autoencoder with adversarial learning for end-to-end text-to-speech.
\newblock In \emph{International Conference on Machine Learning}, pages 5530--5540. PMLR.

\bibitem[{Manocha and Kumar(2022)}]{noresqa}
Pranay Manocha and Anurag Kumar. 2022.
\newblock \href {https://arxiv.org/abs/2206.12285} {Speech quality assessment through mos using non-matching references}.
\newblock \emph{Preprint}, arXiv:2206.12285.

\bibitem[{MediaGEN(2021)}]{dataset_ai_hub}
MediaGEN. 2021.
\newblock Ai-hub multi speaker speech synthesis dataset.
\newblock \url{https://aihub.or.kr/aihubdata/data/view.do?currMenu=115&topMenu=100&aihubDataSe=data&dataSetSn=542}.

\bibitem[{Nekvinda and Dušek(2020)}]{Meta_multilingual}
Tomáš Nekvinda and Ondřej Dušek. 2020.
\newblock \href {https://arxiv.org/abs/2008.00768} {One model, many languages: Meta-learning for multilingual text-to-speech}.
\newblock \emph{Preprint}, arXiv:2008.00768.

\bibitem[{Puchtler et~al.(2021)Puchtler, Wirth, and Peinl}]{dataset_german}
Pascal Puchtler, Johannes Wirth, and Ren{\'e} Peinl. 2021.
\newblock Hui-audio-corpus-german: A high quality tts dataset.
\newblock In \emph{KI 2021: Advances in Artificial Intelligence: 44th German Conference on AI, Virtual Event, September 27--October 1, 2021, Proceedings 44}, pages 204--216. Springer.

\bibitem[{Radford et~al.(2023)Radford, Kim, Xu, Brockman, McLeavey, and Sutskever}]{whisper}
Alec Radford, Jong~Wook Kim, Tao Xu, Greg Brockman, Christine McLeavey, and Ilya Sutskever. 2023.
\newblock Robust speech recognition via large-scale weak supervision.
\newblock In \emph{International Conference on Machine Learning}, pages 28492--28518. PMLR.

\bibitem[{Ravanelli et~al.(2021)Ravanelli, Parcollet, Plantinga, Rouhe, Cornell, Lugosch, Subakan, Dawalatabad, Heba, Zhong, Chou, Yeh, Fu, Liao, Rastorgueva, Grondin, Aris, Na, Gao, Mori, and Bengio}]{speechbrain}
Mirco Ravanelli, Titouan Parcollet, Peter Plantinga, Aku Rouhe, Samuele Cornell, Loren Lugosch, Cem Subakan, Nauman Dawalatabad, Abdelwahab Heba, Jianyuan Zhong, Ju-Chieh Chou, Sung-Lin Yeh, Szu-Wei Fu, Chien-Feng Liao, Elena Rastorgueva, François Grondin, William Aris, Hwidong Na, Yan Gao, Renato~De Mori, and Yoshua Bengio. 2021.
\newblock \href {https://arxiv.org/abs/2106.04624} {{SpeechBrain}: A general-purpose speech toolkit}.
\newblock \emph{Preprint}, arXiv:2106.04624.
\newblock ArXiv:2106.04624.

\bibitem[{Ren et~al.(2021)Ren, Hu, Tan, Qin, Zhao, Zhao, and Liu}]{FastSpeech2}
Yi~Ren, Chenxu Hu, Xu~Tan, Tao Qin, Sheng Zhao, Zhou Zhao, and Tie-Yan Liu. 2021.
\newblock Fastspeech2: Fast and high-quality end-to-end text to speech.
\newblock In \emph{ICLR}.

\bibitem[{Solak(2019)}]{dataset_m-ailabs}
Imdat~Celeste Solak. 2019.
\newblock The m-ailabs speech dataset.
\newblock \url{https://www.caito.de/2019/01/03/the-m-ailabs-speech-dataset/}.

\bibitem[{SPIRE~lab(2023)}]{dataset_limmits24}
India. SPIRE~lab, Indian Institute of Science (IISc)~Bangalore. 2023.
\newblock Multi-speaker, multi-lingual indic tts with voice cloning limmits'24 dataset.
\newblock \url{https://sites.google.com/view/limmits24/dataset/tts-training-data}.

\bibitem[{Tan et~al.(2024)Tan, Chen, Liu, Cong, Zhang, Liu, Wang, Leng, Yi, He et~al.}]{Naturalspeech}
Xu~Tan, Jiawei Chen, Haohe Liu, Jian Cong, Chen Zhang, Yanqing Liu, Xi~Wang, Yichong Leng, Yuanhao Yi, Lei He, et~al. 2024.
\newblock Naturalspeech: End-to-end text-to-speech synthesis with human-level quality.
\newblock \emph{IEEE Transactions on Pattern Analysis and Machine Intelligence}.

\bibitem[{Vaswani et~al.(2017)Vaswani, Shazeer, Parmar, Uszkoreit, Jones, Gomez, Kaiser, and Polosukhin}]{transformers}
Ashish Vaswani, Noam Shazeer, Niki Parmar, Jakob Uszkoreit, Llion Jones, Aidan~N. Gomez, Lukasz Kaiser, and Illia Polosukhin. 2017.
\newblock Attention is all you need.

\bibitem[{Yoon et~al.(2023)Yoon, Kim, Um, Yoon, and Kang}]{SC-CNN}
Hyungchan Yoon, Changhwan Kim, Seyun Um, Hyun-Wook Yoon, and Hong-Goo Kang. 2023.
\newblock \href {https://doi.org/10.1109/LSP.2023.3277786} {Sc-cnn: Effective speaker conditioning method for zero-shot multi-speaker text-to-speech systems}.
\newblock \emph{IEEE Signal Processing Letters}, 30:593--597.

\bibitem[{Zhang et~al.(2019)Zhang, Weiss, Zen, Wu, Chen, Skerry-Ryan, Jia, Rosenberg, and Ramabhadran}]{zhang2019learning}
Yu~Zhang, Ron~J. Weiss, Heiga Zen, Yonghui Wu, Zhifeng Chen, RJ~Skerry-Ryan, Ye~Jia, Andrew Rosenberg, and Bhuvana Ramabhadran. 2019.
\newblock \href {https://arxiv.org/abs/1907.04448} {Learning to speak fluently in a foreign language: Multilingual speech synthesis and cross-language voice cloning}.
\newblock \emph{Preprint}, arXiv:1907.04448.

\end{thebibliography}

\appendix

\section{MOS Prediction Result}
\label{sec:appendix}

\begin{table}[htbp]
\centering
\renewcommand{\arraystretch}{1.4}
\setlength{\tabcolsep}{4pt}
\small
\begin{tabular}{l*{3}{c}}
\hline
 & Baseline & \makecell{Proposed \\ (w/o SAT)} & \makecell{Proposed \\ (with SAT)} \\
\hline
German & 4.34 ± 0.12 & 4.45 ± 0.15 & \textbf{4.45 ± 0.13} \\
French & 4.35 ± 0.23 & 4.36 ± 0.20 & \textbf{4.42 ± 0.13} \\
Italian & 4.24 ± 0.28 & 4.33 ± 0.22 & \textbf{4.40 ± 0.13} \\
Korean & \textbf{4.44 ± 0.07} & 4.43 ± 0.15 & 4.43 ± 0.08 \\
Hindi  & 4.47 ± 0.19 & 4.56 ± 0.26 & \textbf{4.57 ± 0.25} \\
Polish & 4.35 ± 0.24 & \textbf{4.42 ± 0.12} & 4.36 ± 0.14 \\
Russian & 4.31 ± 0.29 & \textbf{4.44 ± 0.09} & 4.43 ± 0.06 \\
Spanish & 4.05 ± 0.44 & 4.27 ± 0.32 & \textbf{4.43 ± 0.08} \\
Ukrainian & 4.40 ± 0.23 & 4.43 ± 0.11 & \textbf{4.43 ± 0.07} \\
\hline
\end{tabular}
\caption{MOS scores of pretrained multi-lingual TTS, with each language evaluated using NORESQA-MOS.}
\label{tab:MOS_Prediction}
\end{table}

To provide a clearer analysis of multilingual TTS, we conducted the MOS prediction task using NORESQA-MOS. As shown in Table~\ref{tab:MOS_Prediction}, the proposed method with SAT outperforms other configurations across most languages, supporting our claim that the approach effectively captures language representations.
\end{document}